\documentclass[conference]{IEEEtran}

\usepackage{graphicx}
\usepackage{balance}

\usepackage{algorithm}
\usepackage{algpseudocode}
\usepackage{subfig}
\usepackage{enumitem}
\usepackage{multirow}
\usepackage{rotating}

\begin{document}

\title{DocLite: A Docker-Based Lightweight Cloud Benchmarking Tool}

\author{%
  Blesson Varghese\\School of EEECS\\Queen's University Belfast, UK\\varghese@qub.ac.uk
  \and Lawan Thamsuhang Subba\\Department of Computer Science\\Aalborg University, Denmark\\lawbba@cs.aau.dk
  \and Long Thai and Adam Barker\\School of Computer Science\\University of St Andrews, UK\\ \{ltt2, adam.barker\}@st-andrews.ac.uk
 }

\maketitle

\begin{abstract}

Existing benchmarking methods are time consuming processes as they typically benchmark the entire Virtual Machine (VM) in order to generate accurate performance data, making them less suitable for real-time analytics. The research in this paper is aimed to surmount the above challenge by presenting \textit{DocLite} - Docker Container-based Lightweight benchmarking tool. DocLite explores lightweight cloud benchmarking methods for rapidly executing benchmarks in near real-time. DocLite is built on the Docker container technology, which allows a user-defined memory size and number of CPU cores of the VM to be benchmarked. The tool incorporates two benchmarking methods - the first referred to as the native method employs containers to benchmark a small portion of the VM and generate performance ranks, and the second uses historic benchmark data along with the native method as a hybrid to generate VM ranks. The proposed methods are evaluated on three use-cases and are observed to be up to 91 times faster than benchmarking the entire VM. In both methods, small containers provide the same quality of rankings as a large container. The native method generates ranks with over 90\% and 86\% accuracy for sequential and parallel execution of an application compared against benchmarking the whole VM. The hybrid method did not improve the quality of the rankings significantly.
\end{abstract}
\begin{IEEEkeywords}
lightweight benchmark; Docker; cloud benchmarking; containers; hybrid benchmark
\end{IEEEkeywords}

\IEEEpeerreviewmaketitle

\section{Introduction}
\label{introduction}
Cloud Virtual Machines (VMs) can be benchmarked to gather, compare and rank performance  \cite{cloudbenchmark-1, cloudbenchmark-2, cloudbenchmark-3, cloudbenchmark-4}. Benchmarking cloud VMs can expose the underlying performance properties across the large number of different cloud providers and instance types. This enables a user to make a more informed choice about which VM(s) would be appropriate for a given computational workload. 

The majority of cloud benchmarking methods are time consuming processes as they benchmark an entire VM and incur significant monetary costs \cite{cloudbenchmark-3}. For example, a VM with 256GB RAM will require a few hours for the memory to be benchmarked. This time consuming process results in significant costs even before an application is deployed on the VM, and means that services which utilise this data (for example, cloud service brokerages \cite{brokerage, cloud} or cloud/cluster management systems) rely on historic data, rather than near real-time data sets. 

We explore methods which execute quickly and can be used in near real-time to collect metrics from cloud providers and VMs, which we have referred to as lightweight cloud benchmarking in this paper. We present DocLite - Docker Container-based Lightweight Benchmarking: a tool to simplify the exploration of lightweight benchmarking methods. Two key research questions are addressed in this paper. Firstly, how fast can lightweight methods execute compared to benchmarking the entire VM? Secondly, how accurate will the generated benchmarks be? 

The Docker\footnote{http://www.docker.com/} container technology \cite{cont-0a, cont-0, cont-1, cont-3}, which allows a user-defined portion (such as memory size and the number of CPU cores) of a VM to be benchmarked is integral to DocLite. The benefit is that containers can be used to benchmark, for example, only 1GB RAM of a VM that has 256GB RAM without benchmarking the entire memory of the VM. This facilitates rapid benchmarking of VMs for use in real-time, which in turn helps to reduce benchmarking costs for the purposes of comparison. 

DocLite organises the benchmark data into four groups, namely memory and process, local communication, computation and storage. A user of DocLite provides as input a set of four weights (ranging from 0 to 5), which indicate how important each of the groups are to the application that needs to be deployed on the cloud. The weights are mapped onto the four benchmark groups and are used to generate a score for ranking the VMs according to performance. Two container-based benchmarking methods are incorporated in DocLite. In the first method, referred to as the \textbf{native method}, containers are used to benchmark a small portion of a VM to generate the performance ranks of VMs. In the second approach, historic benchmark data obtained from benchmarking an entire VM or from previous executions of the native method are used as a \textbf{hybrid method} in order to generate VM ranks. 

Three case study applications are used to validate the benchmarking methods. The experiments highlight that the lightweight methods are up to 91 times faster than benchmarking the entire VM. The rankings generated by lightweight methods are over 90\% and 86\% accurate for sequential and parallel execution of applications when compared to benchmarking the whole VM. The quality of rankings when using the hybrid method is not significantly improved over the native method. 

This paper makes five research contributions. 
Firstly, the development of DocLite, a tool that incorporates container-based cloud benchmarking methods. 
Secondly, the development of lightweight cloud benchmarking methods that can benchmark VMs in near real-time for generating performance ranks. 
Thirdly, an evaluation of the time taken by lightweight benchmarking methods in comparison to methods which benchmark an entire VM.
Fourthly, an evaluation using containers of varying sizes of VM memory against the whole VM. 
Fifthly, an evaluation of the accuracy of the benchmarks generated by DocLite on three case studies.

The remainder of this paper is organised as follows. 
Section \ref{doclite} considers the implementation of DocLite that is developed to explore container-based cloud.  
Section \ref{benchmarking} presents two container-based benchmarking methods incorporated in DocLite.
Section \ref{studies} is an experimental evaluation of the proposed methods using three case study applications. 
Section \ref{relatedwork} considers related research.
Section \ref{conclusions} concludes this paper by reporting future work.  

\section{DocLite Tool}
\label{doclite}
In this section, we consider the building blocks and the system design of DocLite - Docker Container-based Lightweight Benchmarking tool, which is developed to explore container-based benchmarking methods on the cloud. DocLite is available from https://github.com/lawansubba/DoCLite.

\begin{figure*}
	\centering
	\includegraphics[width=0.87\textwidth]{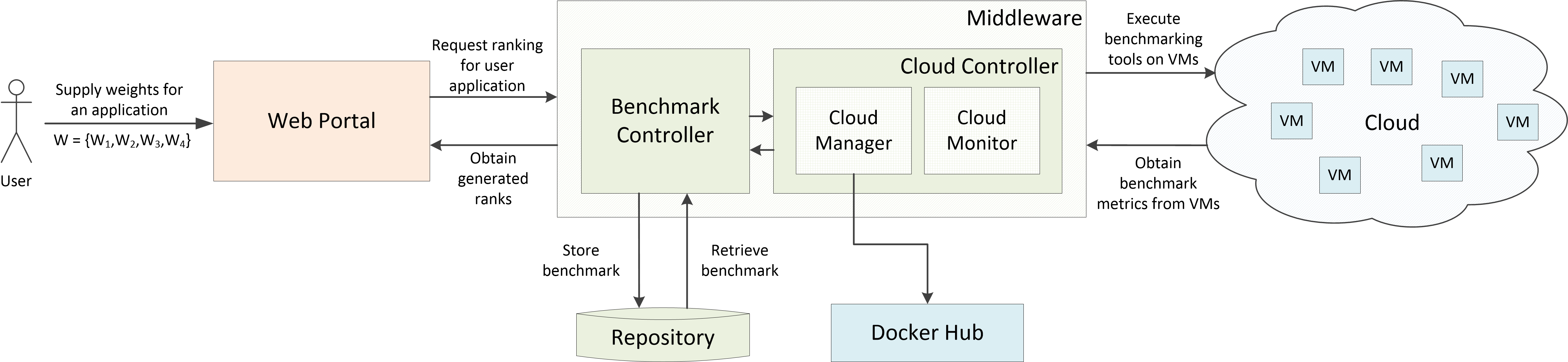}
	\caption{Architecture of DocLite}
	\label{fig:architecture}
\end{figure*}

\subsection{Building Blocks}
\label{implementation:cloudvms}

DocLite is built on (1) the Docker container technology, (2) container-based benchmarking methods, (3) standard benchmarking tools, and (4) a cloud test-bed.

\subsubsection{Docker container technology}
Docker is a portable and lightweight tool that facilitates the execution of distributed applications. It is advantageous in that container images require less storage space and consequentially deployment of containers is quick. Another useful feature of containers is resource isolation - the resources of a VM can be restricted to a specified amount of memory or number of cores (virtual CPUs) for benchmarking. The experiments were performed on 100MB, 500MB and 1000MB of RAM and on a single and on all vCPUs of the VM. In our approach Docker containers are used on top of the VMs and the resulting overhead is negligible as reported by industry experts \cite{foot-2} \cite{foot-3}.

\subsubsection{Container-based benchmarking methods}
DocLite incorporates two container-based benchmarking methods which will be considered in the next section. In the first method, referred to as the native method, a user-defined portion (such as memory size and the number of CPU cores) of a VM can be rapidly benchmarked thereby obtaining benchmark data in real-time. DocLite then organises the benchmark data into four groups, namely memory and process, local communication, computation and storage. A DocLite user provides as input a set of four weights (ranging from 0 to 5), which indicate how important each of the groups are to the application that needs to be deployed on the cloud. The weights are mapped onto the four benchmark groups to generate a score for ranking the VMs according to performance. In the second method, historic benchmark data is used along with the native method as a hybrid in order to generate VM ranks.

\subsubsection{Benchmarking tools}
The benchmarking tool employed in this paper is \texttt{lmbench} \cite{lmbench-1}. It was selected since (i) it is a single tool and can be easily deployed on the cloud, (ii) it provides a wide variety of benchmarks related to memory and process, computation, local communication and file related operations that capture the performance characteristics of the VM, and (iii) it has been previously employed for modelling the performance of cloud VMs \cite{cloudbenchmark-1, cloudbenchmark-3, cloudbenchmark-100}. In addition to the execution of \texttt{lmbench}, other benchmarking tools can be executed independently or in any preferred combination on DocLite.

\subsubsection{Cloud test-bed}
The cloud test-bed used in this research is the Amazon Web Services (AWS) Elastic Compute Cloud (EC2)\footnote{http://aws.amazon.com/ec2/previous-generation/}. The previous generation VMs, shown in Table \ref{table1}, which have varying performance and have become popular in the scientific community due to their longevity are chosen. 

\begin{table}[ht]
	\caption{Amazon EC2 VMs employed for benchmarking}
	\label{table1}
\begin{center}
	\begin{tabular}{c c p{0.7cm} p{2.3cm} p{0.9cm}}
		\hline	
		\textbf{VM Type}	&	\textbf{vCPUs}	&	\textbf{Memory (GiB)}	&	\textbf{Processor}	& \textbf{Clock (GHz)}	\\
		\hline	
		\hline	
		\texttt{m1.xlarge}	&	4	&	15.0	&	Intel Xeon E5-2650	&	2.00\\
		\texttt{m2.xlarge}	&	2	&	17.1	&	Intel Xeon E5-2665	&	2.40\\
		\texttt{m2.2xlarge}	&	4	&	34.2	&	Intel Xeon E5-2665	&	2.40\\
		\texttt{m2.4xlarge}	&	8	&	68.4	&	Intel Xeon E5-2665	&	2.40\\
		\texttt{m3.xlarge}	&	4	&	15.0	&	Intel Xeon E5-2670	&	2.60\\
		\texttt{m3.2xlarge}	&	8	&	30.0	&	Intel Xeon E5-2670	&	2.60\\
		\texttt{cr1.8xlarge}	&	32	&	244.0	&	Intel Xeon E5-2670	&	2.60\\		
		\texttt{cc2.8xlarge}	&	32	&	60.5	&	Intel Xeon X5570	&	2.93\\
		\texttt{hi1.4xlarge}	&	16	&	60.5	&	Intel Xeon E5620	&	2.40\\
		\texttt{hs1.8xlarge}&	16	&	117.0	&	Intel Xeon E5-2650	&	2.00\\
		\hline
	\end{tabular}
	\end{center}
\end{table}

\subsection{System Design}
\label{implementation:architecture}

DocLite has three main components, namely (1) a web portal, (2) middleware and (3) a benchmark repository as shown in Figure \ref{fig:architecture}.

\subsubsection{Web Portal}
\begin{figure}
	\centering
	\includegraphics[width=0.47\textwidth]{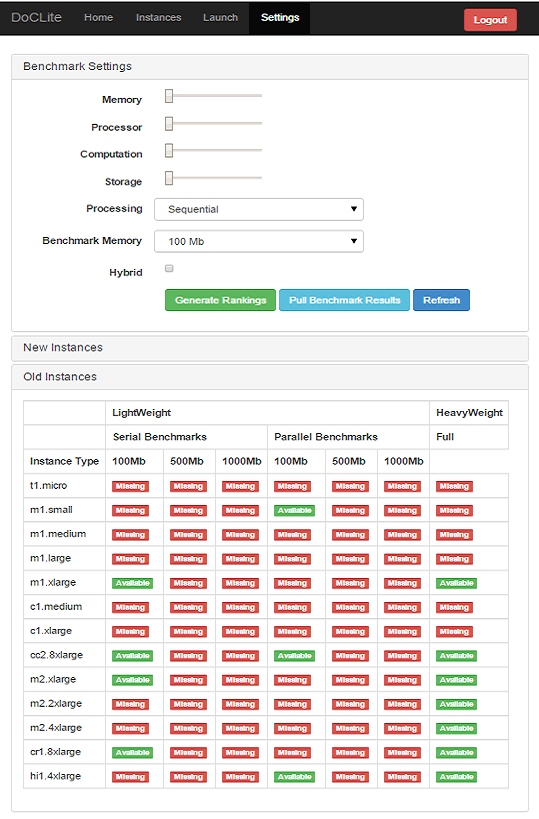}
	\caption{DocLite screenshot; top shows the sliders for providing weights, bottom shows list of Amazon instances benchmarked (`\textit{Available}' indicates benchmarking data is in the repository, and `\textit{Missing}' shows instances that need to be benchmarked)}
	\label{fig:doclitescreenshot1}
\end{figure}

This user facing component is developed using MVC.NET and Bootstrap. A user provides as input a set of four weights $W=\{W_1, W_2, W_3, W_4\}$ that characterises the application to be deployed on the cloud; the amount of memory and number of cores to be benchmarked along with preferences of whether the benchmark needs to be executed sequentially or in parallel. The portal is also responsible for displaying the status of the cloud VMs that are used and the ranks generated from the benchmarks. A screenshot of DocLite is shown in Figure \ref{fig:doclitescreenshot1}.

\subsubsection{Middleware}
This component comprises a Benchmark Controller and a Cloud Controller. The Benchmark Controller (i) incorporates the algorithms for container-based benchmarking methods considered in Section \ref{benchmarking}, (ii) pulls benchmark data from the repository for grouping and normalising the data, and (iii) generates the score for each VM based on the weights provided by the user. 

The Cloud Controller comprises of a Cloud Manager and a Cloud Monitor. The manager initiates cloud VMs and maintains them by executing the appropriate scripts for installing necessary packages and setting up Docker on the VM. The Docker images that are used are retrieved from the Docker Hub\footnote{https://hub.docker.com/} by the manager. The benchmarked data is deposited by the manager into the repository.

The monitor keeps track of the benchmarks that have started on the cloud VMs and reports the status of the VM to the portal. Monitoring is important to the Benchmark Controller to retrieve data from the repository after benchmarking.

\subsubsection{Benchmark Repository}
The benchmark data obtained from the cloud is stored in a repository used by the Benchmark Controller for generating scores. Both historic and current benchmark data are stored in the repository. If the native method is chosen, then the current benchmark data is used, where as if the hybrid method is chosen, then the historic data is used along with the current data. 

\section{Container-Based Benchmarking Methods}
\label{benchmarking}
In this section, two container-based cloud benchmarking methods are proposed and presented. The first is referred to as a native method that employs Docker container technology to benchmark cloud VMs in real-time. The second method, combines the use of benchmarks from the first along with historic benchmarks generated by benchmarking the entire VM as a hybrid. The aim of both methods is to generate a ranking of cloud VMs based on performance. The benefit of using containers is that the amount of VM resources benchmarked, such as memory size and the number of CPU cores, can be limited. This benefit is leveraged such that only a small amount of resources available to a VM are benchmarked in comparison to benchmarking the entire resource of a VM. 

A user can opt for either the native or hybrid method. A set of user defined weights, $W$ (considered later in this section), and historic benchmark data, $HB$, obtained from benchmarking the entire VM or from previous executions of the lightweight methods can be provided as input. The lightweight benchmarks are obtained from \texttt{Obtain-Benchmark}, considered in Algorithm \ref{algorithm2}, which is represented as $B$. \texttt{Native-Method} (Algorithm \ref{algorithm3}) takes as input $W$ and $B$ and \texttt{Hybrid-Method} (Algorithm \ref{algorithm4}) takes $HB$ as additional input.

Algorithm \ref{algorithm2} gathers benchmarks from different cloud VMs. 
Consider there are $i = 1, 2, \cdots , m$ different VM types, and a VM of type $i$ is represented as $vm_{i}$ as shown in \textit{Line 2}. 
A container $c_{i}$ is created on each VM in \textit{Line 3}. In the context of cloud benchmarking, containers are used to facilitate benchmarking on different types of VMs by restricting the amount of resources used.

\begin{algorithm} 
	\caption{Obtain cloud benchmarks}
	\label{algorithm2}
	\begin{algorithmic}[1]
		\Procedure{Obtain\textendash Benchmark}{$mem$, $CPU\_cores$}{}
			\For{each virtual machine $vm_{i} \in VM$ }
				\State Create container $c_{i}$ of $mem$ size and $CPU\_cores$ on $vm_{i}$
				\State Execute standard benchmark tool on $c_{i}$
				\State Store benchmarks as $B$
		\EndFor
	\EndProcedure
	\end{algorithmic}
\end{algorithm}

In \textit{Line 4}, standard tools are executed using the container $c_{i}$ to benchmark a portion of the VM. In this research, the VMs shown in Table \ref{table1} were benchmarked using DocLite by executing lmbench.
The latency and bandwidth information for a wide range of memory and process, computation, local communication and file related operations are collected. In \textit{Line 5}, the benchmarks obtained for each $vm_{i}$ are stored in a file, $B$, for use by Algorithm \ref{algorithm3} and Algorithm \ref{algorithm4}. 

\begin{figure*}[ht]
\centering
	\subfloat[Main memory latency]{\label{figure3a}\includegraphics[width=0.32\textwidth]{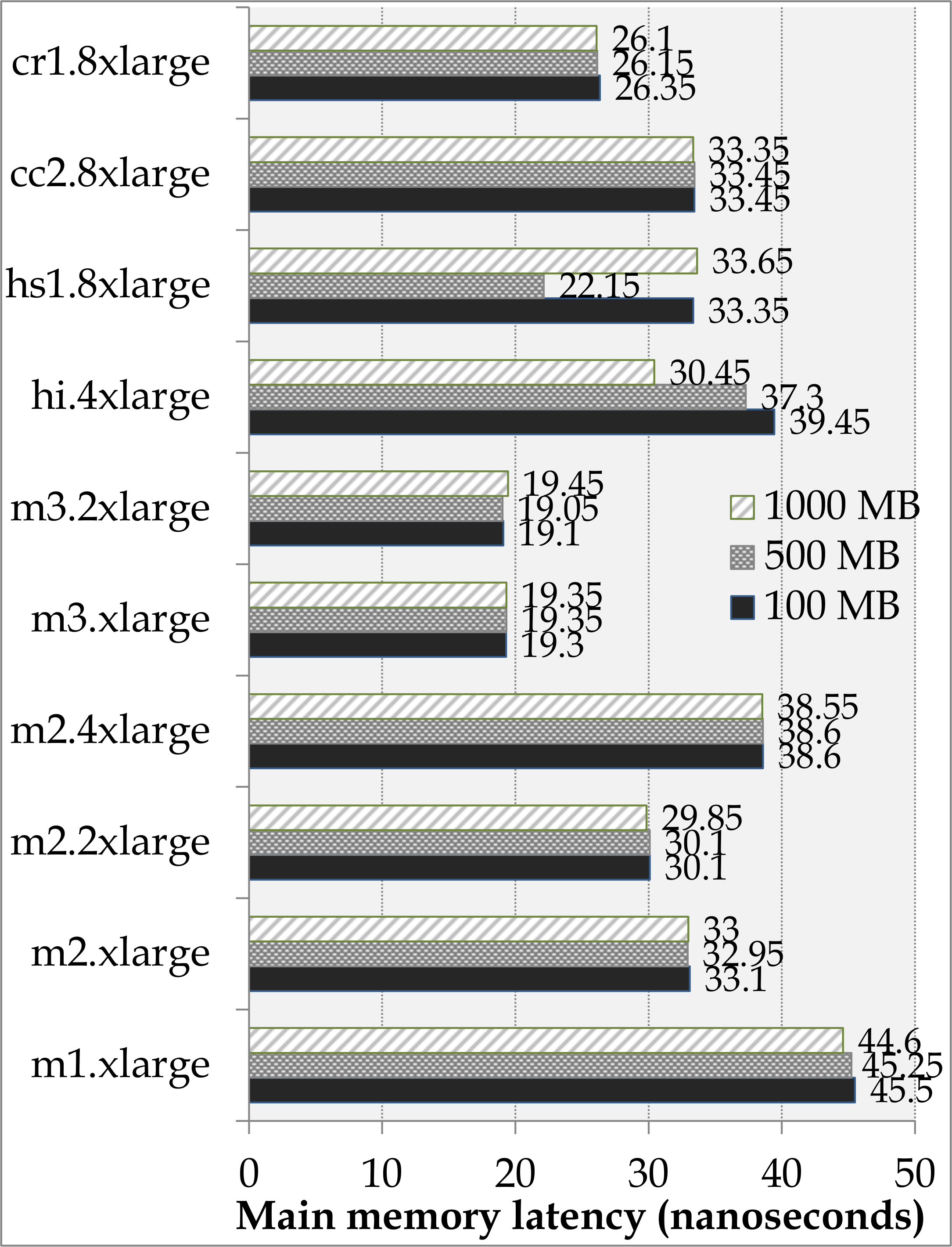}} \hfill
	\subfloat[Float division operation latency]{\label{figure3b}\includegraphics[width=0.32\textwidth]{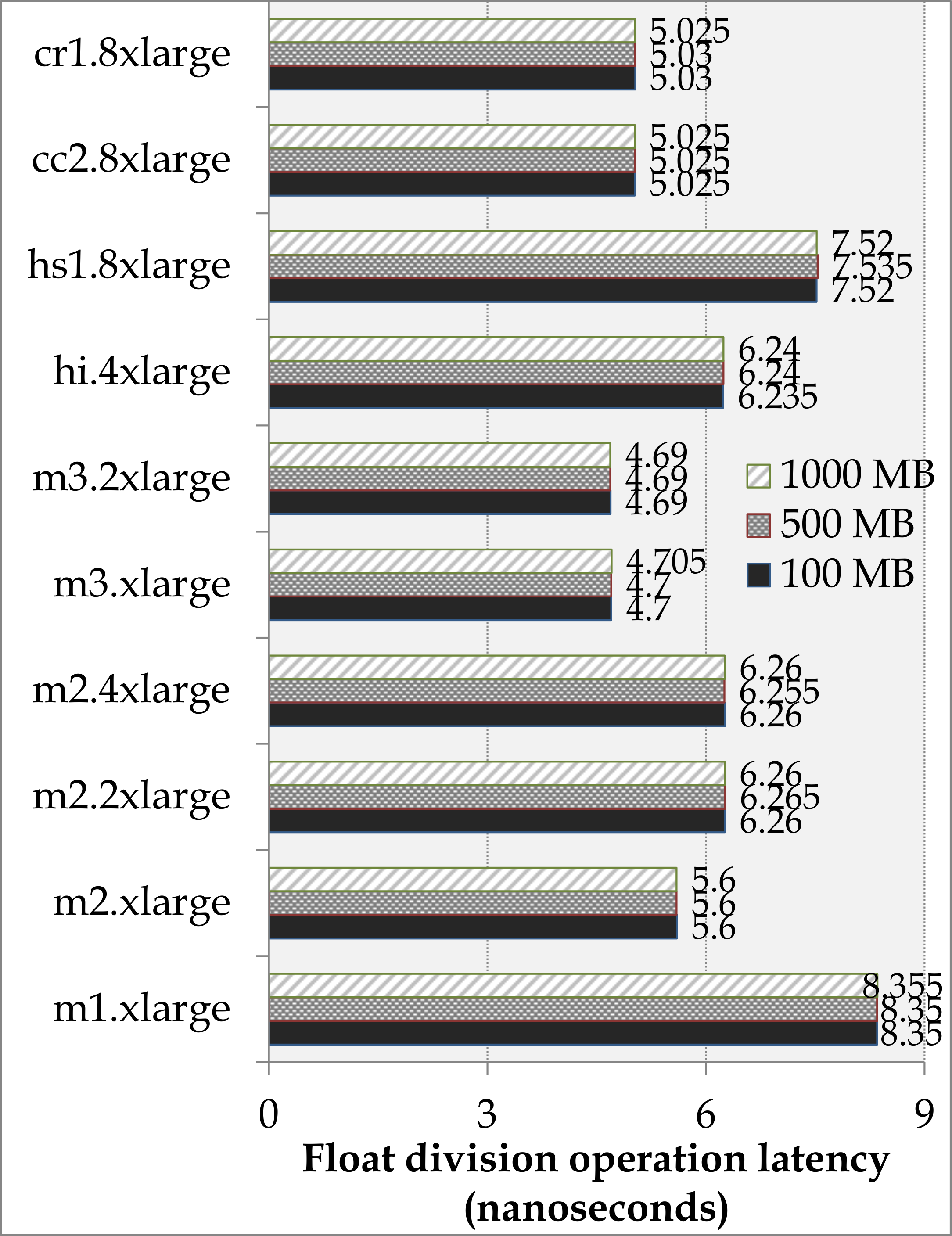}} \hfill
	\subfloat[Memory read bandwidth]{\label{figure3c}\includegraphics[width=0.32\textwidth]{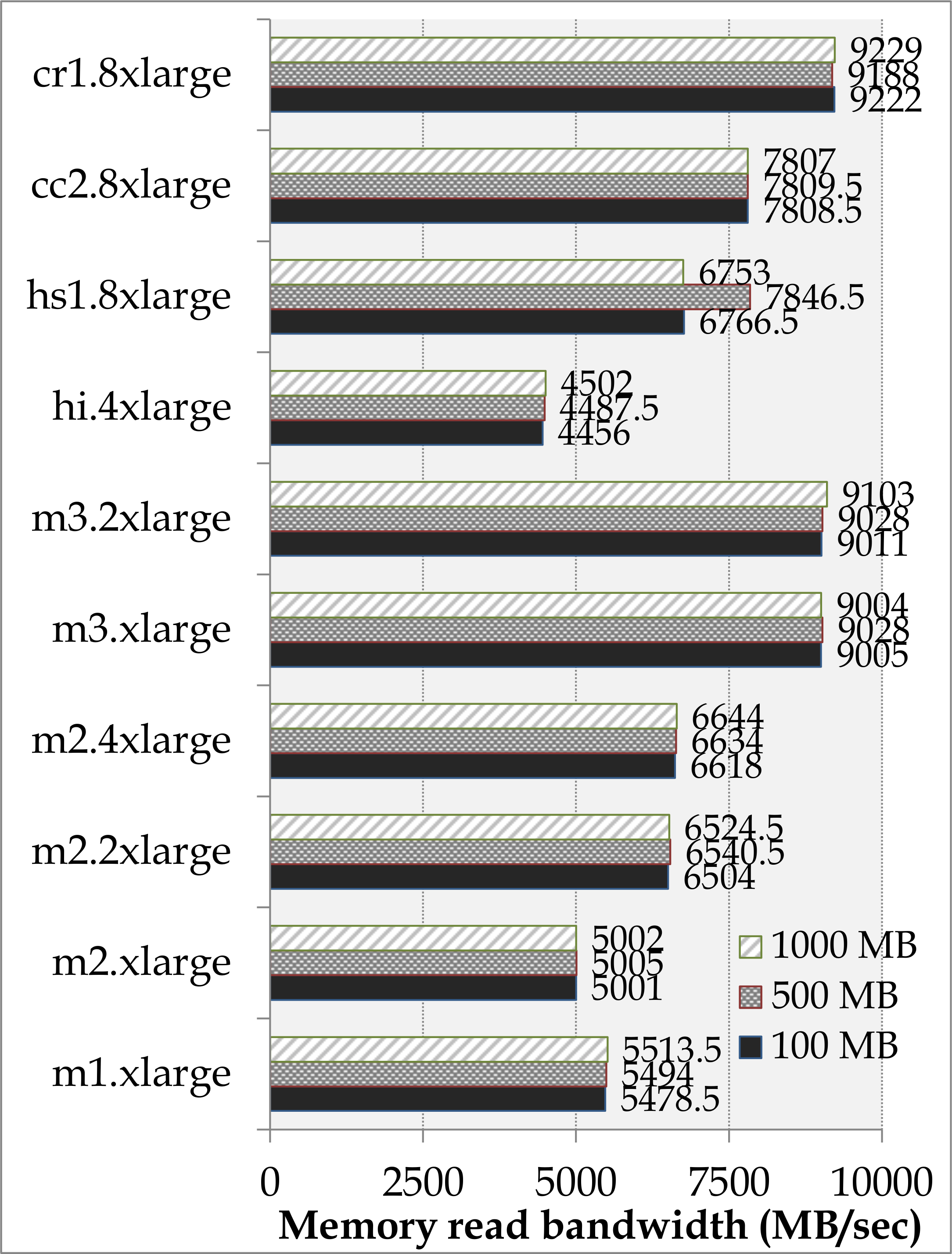}}
\caption{Sample lmbench benchmarks obtained from DocLite for 100MB, 500MB and 1000MB containers}
\label{figure3}
\end{figure*}

Benchmarks for over fifty attributes related to memory and process, local communication, computation, and storage were obtained using containers of 100MB, 500MB and 1000MB. It is not within the scope of this paper to present all benchmarks. Therefore, a sample of three benchmarks is presented as shown in Figure \ref{figure3}.

Figure \ref{figure3a} shows the main memory latency for all VMs. It is immediately inferred that with the exception of \texttt{hs1.8xlarge} and \texttt{hi1.4xlarge} the main memory latencies are comparable for different container sizes. The exceptions are artefacts of measurements over networks. The best main memory performance is for the \texttt{m3} instances. Figure \ref{figure3b} shows the latency for a float division operation on each VM. Again, similar results are obtained for different container sizes. The bandwidth of memory read operations on all VMs are shown in Figure \ref{figure3c}. Maximum memory bandwidth is available on \texttt{cr1.8xlarge}. 

The key observation from the above three samples (also observed in the entire benchmarked data) is that there is a minimal difference on average of less than 2\% between the data points when a container of 100MB, 500MB or 1000MB is used. Given this small difference for different container sizes and the time taken to benchmark a VM using a small container is lower than a larger container, we hypothesise that (i) the benchmarking methods incorporated in DocLite can be employed in real-time, and (ii) VM rankings generated using the lightweight benchmarking methods will be comparable to methods that benchmark the entire VM. This hypothesis will be evaluated in Section \ref{studies} using three case study applications.

\subsection{Native Method}
\label{lightweight}
The native method for generating ranks of VMs based on performance is shown in Algorithm \ref{algorithm3}. Historic benchmark data is not employed in this method. The benchmark data obtained from using containers, $B$, is used as shown in \textit{Line 1}. 

\begin{algorithm} 
	\caption{Cloud ranking using native method}
	\label{algorithm3}
	\begin{algorithmic}[1]
		\Procedure{Native\textendash Method}{$W$, $B$}{}
			\State From $B$, organise benchmarks into groups, $G$
			\State Normalise groups, $\bar{G}$
			\State Score $VM$ using $\bar{G}.W$
			\State Generate performance ranks $R_p$ 
		\EndProcedure
	\end{algorithmic}
\end{algorithm}

Consider there are $j = 1, 2, \cdots , n$ attributes of a VM that are benchmarked, and $r_{i,j}$ is the value associated with each $j^{th}$ attribute on the $i^{th}$ VM. In \textit{Line 2}, the attributes can be grouped as $G_{i, k} = \{r_{i, 1}, r_{i, 2}, \cdots \}$, where $i = 1, 2, \cdots m$, $k = 1, 2, \cdots , p$, and $p$ is the number of attribute groups. In this paper, four attribute groups are considered.

Firstly, \textit{Memory and Process Group}, denoted as $G_{1}$ captures the performance and latencies of the processor. Main memory and random memory latencies of the VMs and L1 and L2 cache latencies are considered in this group.

Secondly, \textit{Local Communication Group} in which the bandwidth of both memory communications and interprocess communications are captured under the local communication group, denoted as $G_{2}$. Memory communication metrics, namely the rate (MB/sec) at which data can be read from and written to memory, and interprocess communication metrics, namely the rate of data transfer between Unix pipes, AF\_Unix sockets and TCP are considered in this group.

Thirdly, \textit{Computation Group}, denoted as $G_{3}$ which captures the performance of integer, float and double operations such as addition, multiplication and division and modulus. 

Fourthly, \textit{Storage Group} in which the file I/O related attributes are grouped together and denoted as $G_{4}$. Sequential create, read and delete and random create, read and delete operations are considered in this group.

In \textit{Line 3}, the attributes of each group are normalised as $\bar{r}_{i, j} = \frac{r_{i, j} - \mu_{j}}{\sigma_{j}}$, where $\mu_j$ is the mean value of attribute $r_{i, j}$ over $m$ VMs and $\sigma_j$ is the standard deviation of the attribute $r_{i, j}$ over $m$ VMs. The normalised groups are denoted as $\bar{G}_{i, k} = \{\bar{r}_{i, 1}, \bar{r}_{i, 2}, \cdots \}$, where $i = 1, 2, \cdots m$, $k = 1, 2, \cdots , p$, and $p$ is the number of groups.

One input to Algorithm \ref{algorithm3} is $W$, which is a set of weights that correspond to each group (for the four groups $G_{1}, G_{2}, G_{3}, G_{4}$, the weights are $W=\{W_{1}, W_{2}, W_{3}, W_{4}\}$). For a given application, a few groups may be more important than the others. For example, if there are a large number of file read and write operations in a simulation, the Storage group represented as $G_{4}$ is important. The weights are provided by the user based on domain expertise and the understanding of the importance of each group to the application. Each weight, $W_{k}$, where $k=1, 2, 3, 4$ takes value between 0 and 5, where 0 indicates that the group is not relevant to the application and 5 indicates the importance of the group for the application. 

In \textit{Line 4}, each VM is scored as $S_{i} = \bar{G}_{i, k}.W_{k}$. In \textit{Line 5}, the scores are ordered in a descending order for generating $Rp_{i}$ which is the ranking of VMs based on performance. 

\subsection{Hybrid Method}
\label{hybrid}

The hybrid method employs benchmarks obtained in real-time, $B$, and historic benchmarks, $HB$, obtained either by executing the benchmark tool on an entire VM or a previous execution of the native method as shown in Algorithm \ref{algorithm4} which is shown in \textit{Line 1}. This method accounts for past and current performance of a VM for generating ranks.

\begin{algorithm}[H]
	\caption{Cloud ranking using hybrid method}
	\label{algorithm4}
	\begin{algorithmic}[1]
		\Procedure{Hybrid\textendash Method}{$W$, $B$, $HB$}{}
			\State From $B$, organise benchmarks into groups, $G$
			\State Normalise groups, $\bar{G}$
			\State From $HB$, organise historic benchmarks into groups, $HG$
			\State Normalise groups, $\bar{HG}$
			\State Score $VM$ using $\bar{G}.W + \bar{HG}.W$
			\State Generate performance ranks $R_p$ 
		\EndProcedure
	\end{algorithmic}
\end{algorithm}

Grouping attributes and normalising them in \textit{Lines 2-3} are similar to those followed in Algorithm \ref{algorithm3}. The four groups used in the native method are used here and the attributes are normalised using the mean and standard deviation values of each attribute.

When historic benchmarks are used, the method for grouping the attributes and normalising them are similar to what was followed previously. In \textit{Line 4}, the attributes from historic benchmark data, $hr$, can be grouped as $HG_{i, k} = \{hr_{i, 1}, hr_{i, 2}, \cdots \}$, where $i = 1, 2, \cdots m$ for $m$ VM types, $k = 1, 2, \cdots , p$, and $p$ is the number of groups. Four groups, $HG_{1}$, $HG_{2}$, $HG_{3}$ and $HG_{4}$, are obtained. 

In \textit{Line 5}, the attributes of each group are normalised as $\bar{hr}_{i, j} = \frac{hr_{i, j} - h\mu_{j}}{h\sigma_{j}}$, where $h\mu_j$ is the mean value of attribute $hr_{i, j}$ over $m$ VMs and $h\sigma_j$ is the standard deviation of the attribute $hr_{i, j}$ over $m$ VMs. The normalised groups are denoted as $\bar{HG}_{i, k} = \{\bar{hr}_{i, 1}, \bar{hr}_{i, 2}, \cdots \}$, where $i = 1, 2, \cdots m$, $k = 1, 2, \cdots , p$, and $p$ is the number of groups.

The set of weights supplied to the hybrid method are same as the native method. Based on the weights each VM is scored as $S_{i} = \bar{G}_{i, k}.W_{k} + \bar{HG}_{i, k}.W_{k}$ in \textit{Line 6}. The scores take the most current and historic benchmarks into account and are ordered in descending order for generating $Rp_{i}$ in \textit{Line 7} which is the performance ranking of the VMs.

Both the native and hybrid methods are incorporated in DocLite and they are further evaluated against real world applications in the next section.

\section{Experimental Studies}
\label{studies}
In this section, three case study applications are used to evaluate the benchmarking methods. The initial hypotheses of this research are: (i) lightweight benchmarking can be employed in real-time, and (ii) VM rankings generated from lightweight benchmarking methods will be comparable to the rankings obtained from benchmarking the entire VM. The evaluation to validate the hypotheses firstly compares the time taken to execute the benchmarks using containers on the VMs and on the entire VM. Secondly, the validation compares VM rankings generated by 
the native and hybrid lightweight benchmarking methods, we refer to benchmarked ranks, and the ranks obtained by execution on the entire VM, referred to as the empirical ranks. 


\begin{table*}
\centering
\renewcommand{\arraystretch}{1.5}%
\begin{tabular}{|p{0.1cm}|p{0.8cm}|c|c|c|c|c|c|c|c|c|c|}
\hline
\multicolumn{2}{|c|}{\multirow{3}{*}{Size}} & \multicolumn{10}{|c|}{Average time (minutes)}\\
\cline{3-12}
 \multicolumn{2}{|c|}{} & m1.xlarge	&	m2.xlarge	&	m2.2xlarge	&	m2.4xlarge	&	m3.xlarge	&	m3.2xlarge	&	hi1.4xlarge	&	hs1.4xlarge	&	cc2.8xlarge	&	cr1.8xlarge\\
\hline
\hline
\multirow{3}{*}{\rotatebox[origin=c]{90}{Container}} & 100MB	&	8	&	8	&	9	&	8	&	7	&	7	&	8	&	9	&	8	&	8\\
\cline{2-12}
& 500MB &	13	&	12	&	13	&	12	&	11	&	11	&	14	&	12	&	13	&	12\\
\cline{2-12}
& 1000MB	&	19	&	19	&	18	&	21	&	15	&	16	&	18	&	17	&	18	&	18\\
\hline
\multicolumn{2}{|c|}{Entire VM}		&	152	&	142	&	293	&	598	&	135	&	152	&	453	&	822	&	341	&	618\\
\hline
\end{tabular}
\caption{Average time for executing benchmarks using 100MB, 500MB and 1000MB containers and on the whole VM}
\label{evaluation:feasibility}
\end{table*}

\subsection{Case Study Applications}
\label{studies:casestudy}
Three applications are chosen to evaluate the benchmarking methods. The first case study is a molecular dynamics simulation of a system comprising 10,000 particles in a three dimensional space used by theoretical physicists \cite{md-1}. The simulation solves differential equations to model particles for different time steps. The simulation is memory intensive with numerous read and write operations and computationally intensive requiring a large number of float operations. Local communication between processes are less relevant and the application does not require file operations. 

The second case study is a risk simulation that generates probable maximum losses due to catastrophic events \cite{risk-1}. The simulation considers over a million alternate views of a given year and a number of financial terms to estimate losses. The simulation is memory intensive with numerous read and write operations and at the same time computationally intensive requiring a large number of float operations to be performed both to compute the risk metrics. The local communication between processes are less relevant and the application does not require file operations.

The third case study is a block triagonal solver, which is a NASA Parallel Benchmark (NPB), version 3.3.1
\footnote{https://www.nas.nasa.gov/publications/npb.html}
\cite{npb-1}. This mathematical solver is used on a grid size of $162 \times 162 \times 162$ for 200 iterations. The solver is numerically intensive and memory and processor related operations are relevant, but does not take precedence over computations. Local communications and file operations have little effect on the solver. 

\subsection{Evaluation}
\label{studies:evaluation}

The aims of the experimental evaluation are to address two important research questions related to lightweight benchmarking. They are: 1) how fast can lightweight benchmarking execute compared to benchmarking the entire VM? and 2) how accurate will the benchmarked data generated from lightweight methods be?

\subsubsection{Execution Time of Benchmarks}
\label{eval:time}
The first question related to speed is addressed by demonstrating the feasibility of the proposed lightweight benchmarking methods in real-time on the cloud. For this, the time taken to execute the lightweight benchmarking methods are compared against the time taken to benchmark the entire VM as shown in Table \ref{evaluation:feasibility}. On an average the 100 MB, 500 MB, and 1000 MB containers take 8 minutes, 13 minutes and 18 minutes to complete benchmarking on all the VMs. Benchmarking the whole VM takes up to 822 minutes for \texttt{hs1.4xlarge}.  
It is immediately evident that container-based benchmarking is between 19-91 times faster than the benchmarking the entire VM.  

\subsubsection{Accuracy of Benchmarks}
\label{evaluation:empiricalanalysis}
The second question related to accuracy is addressed by evaluating the lightweight methods against three real-world case study applications. For this, ranks obtained from DocLite by benchmarking the application based on a user's input weights that describe an application are compared against actual ranks of VMs when the application is executed on the VM. The following five steps are used to evaluate the accuracy of the benchmarks:

\begin{figure*}[ht]
\centering
	\subfloat[Case study 1 - sequential]{\label{figure1-1}\includegraphics[width=0.32\textwidth]{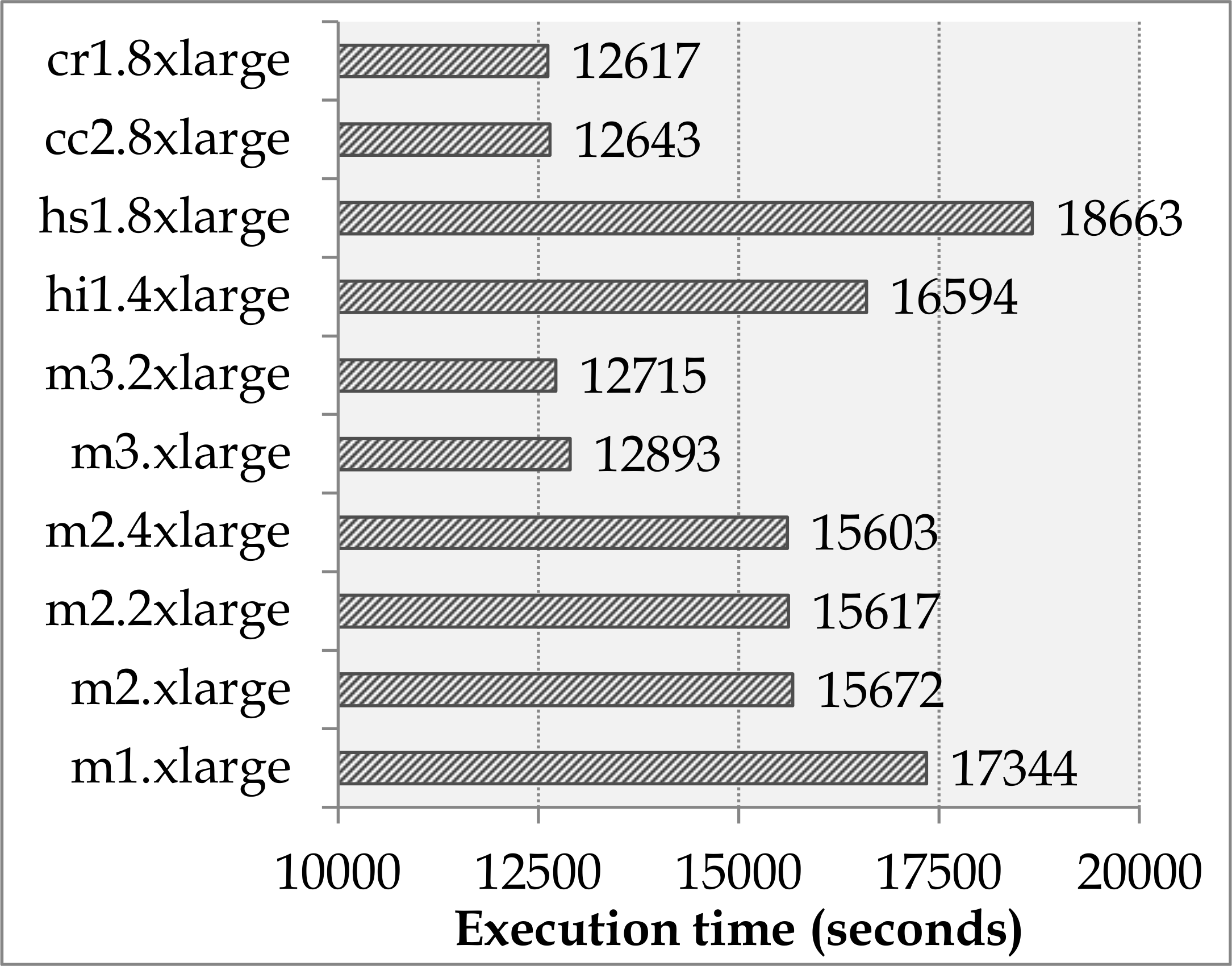}} \hfill
	\subfloat[Case study 2 - sequential]{\label{figure1-2}\includegraphics[width=0.32\textwidth]{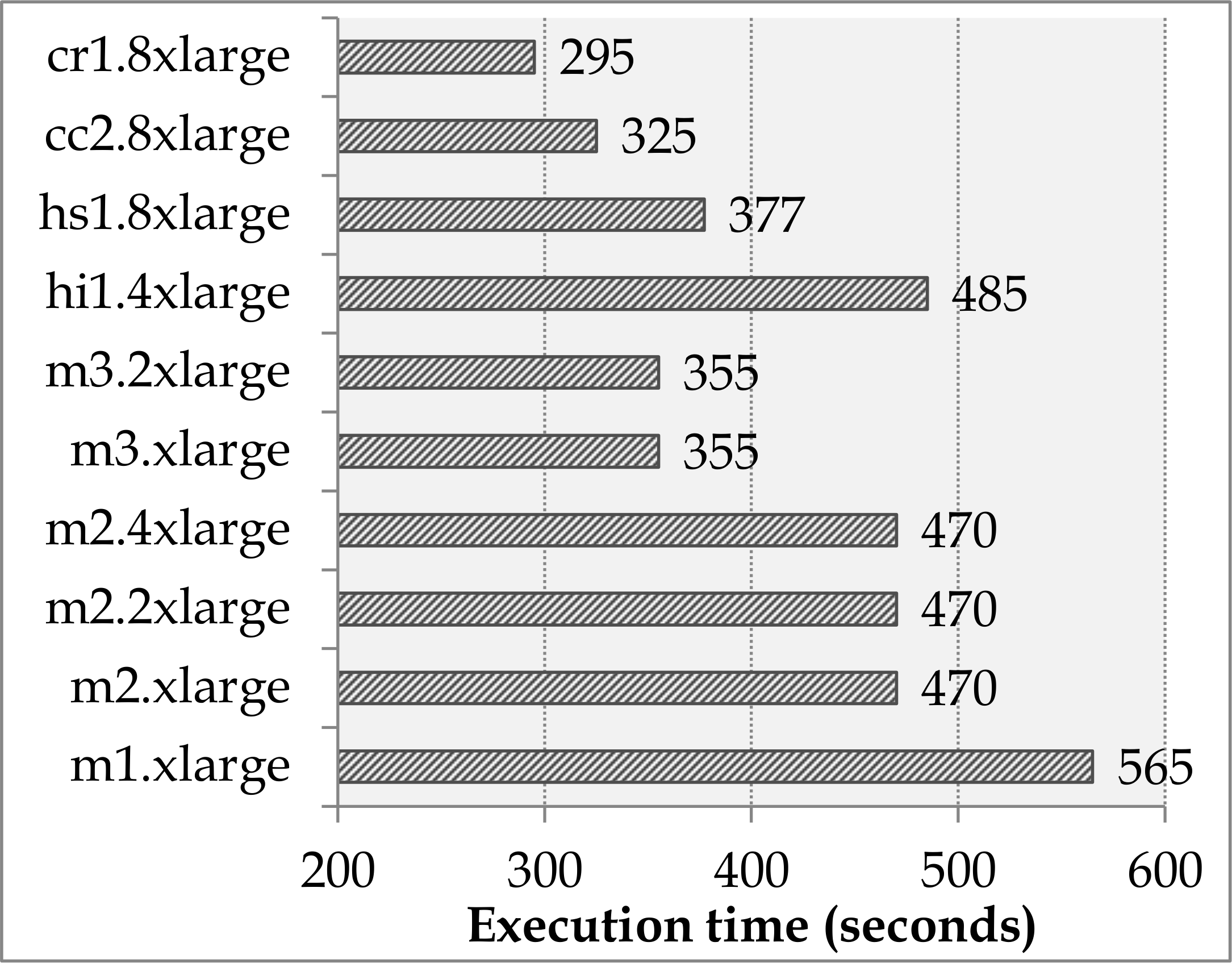}} \hfill
	\subfloat[Case study 3 - sequential]{\label{figure1-3}\includegraphics[width=0.32\textwidth]{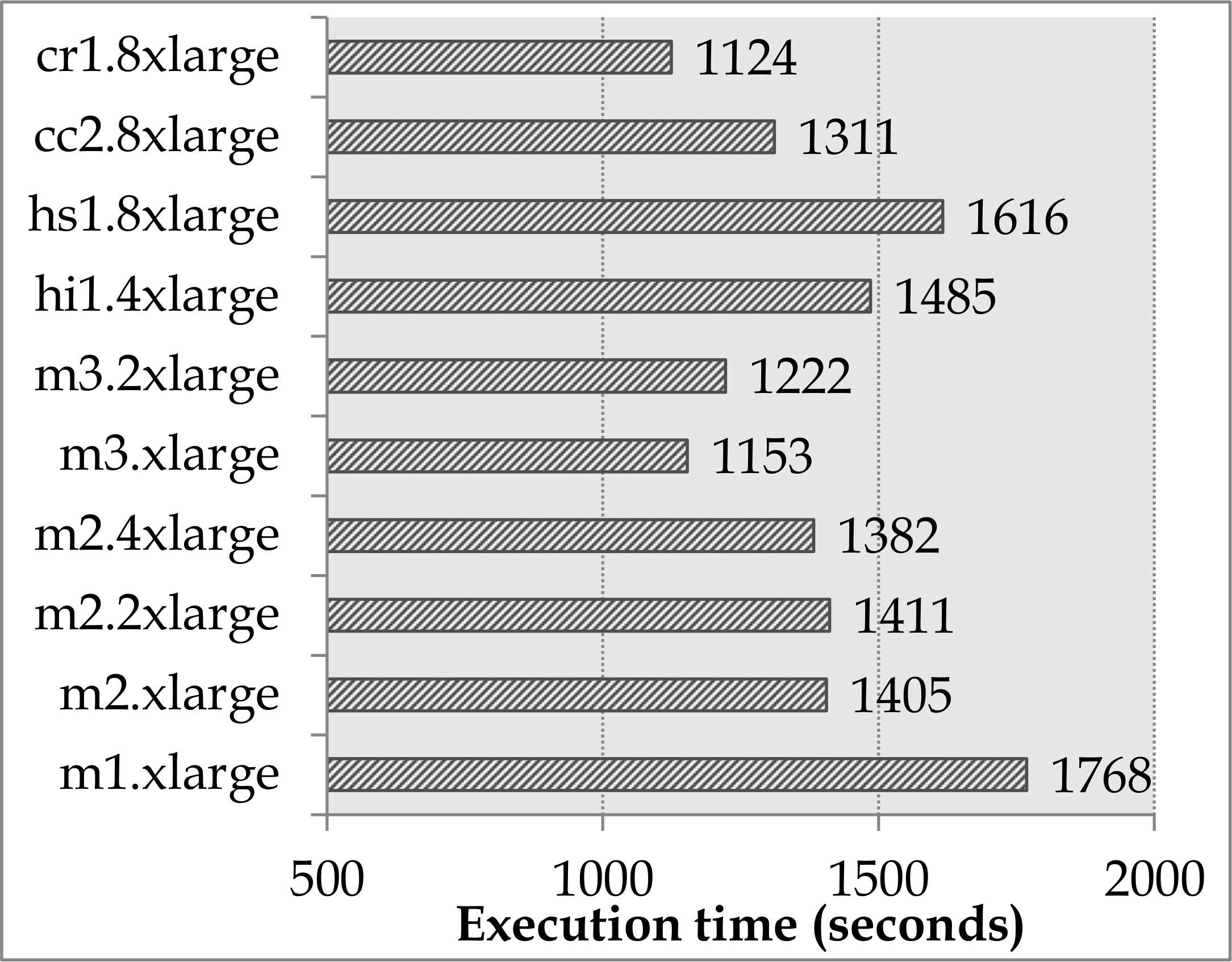}}\\
	\subfloat[Case study 1 - parallel]{\label{figure2-1}\includegraphics[width=0.32\textwidth]{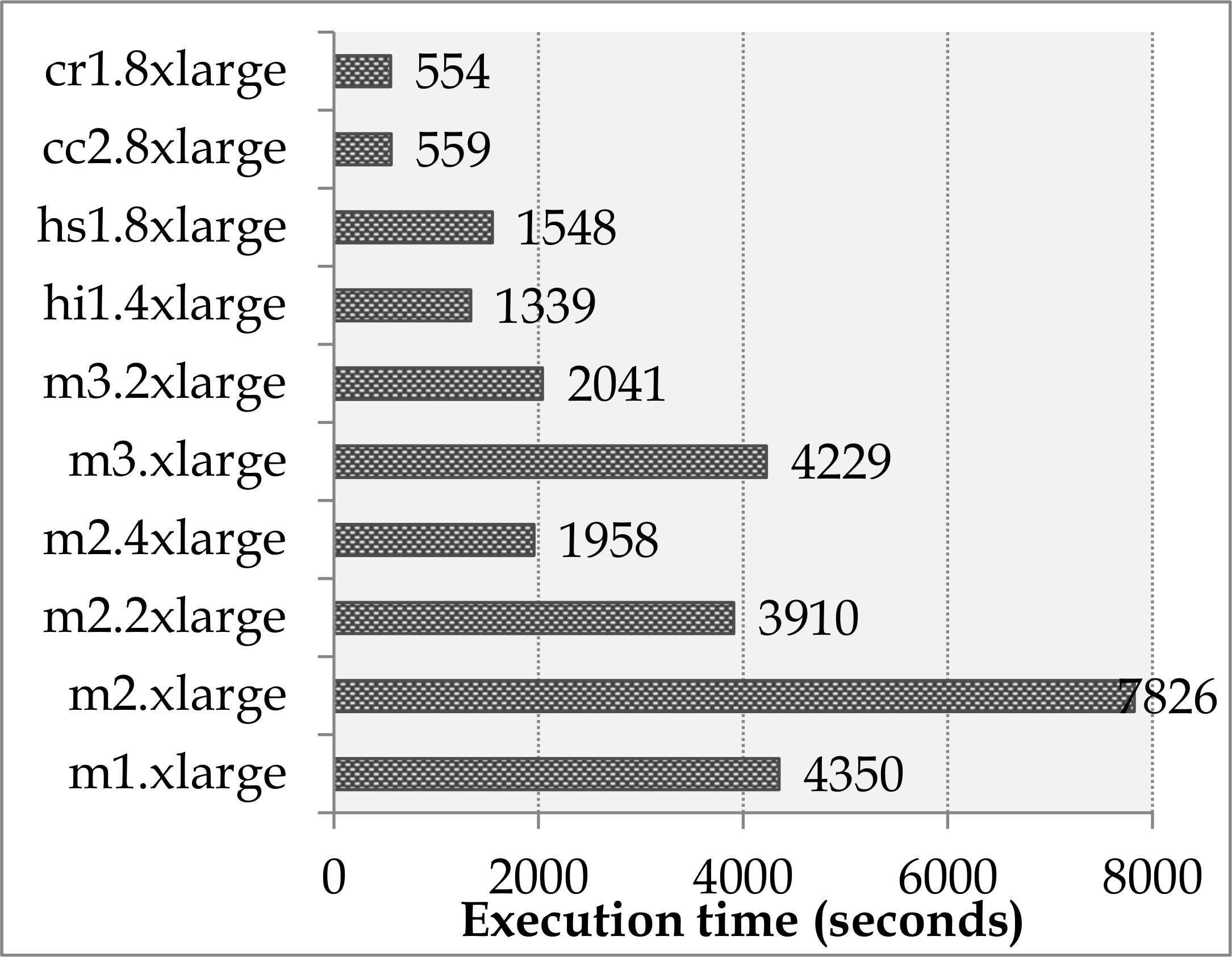}} \hfill
	\subfloat[Case study 2 - parallel]{\label{figure2-2}\includegraphics[width=0.32\textwidth]{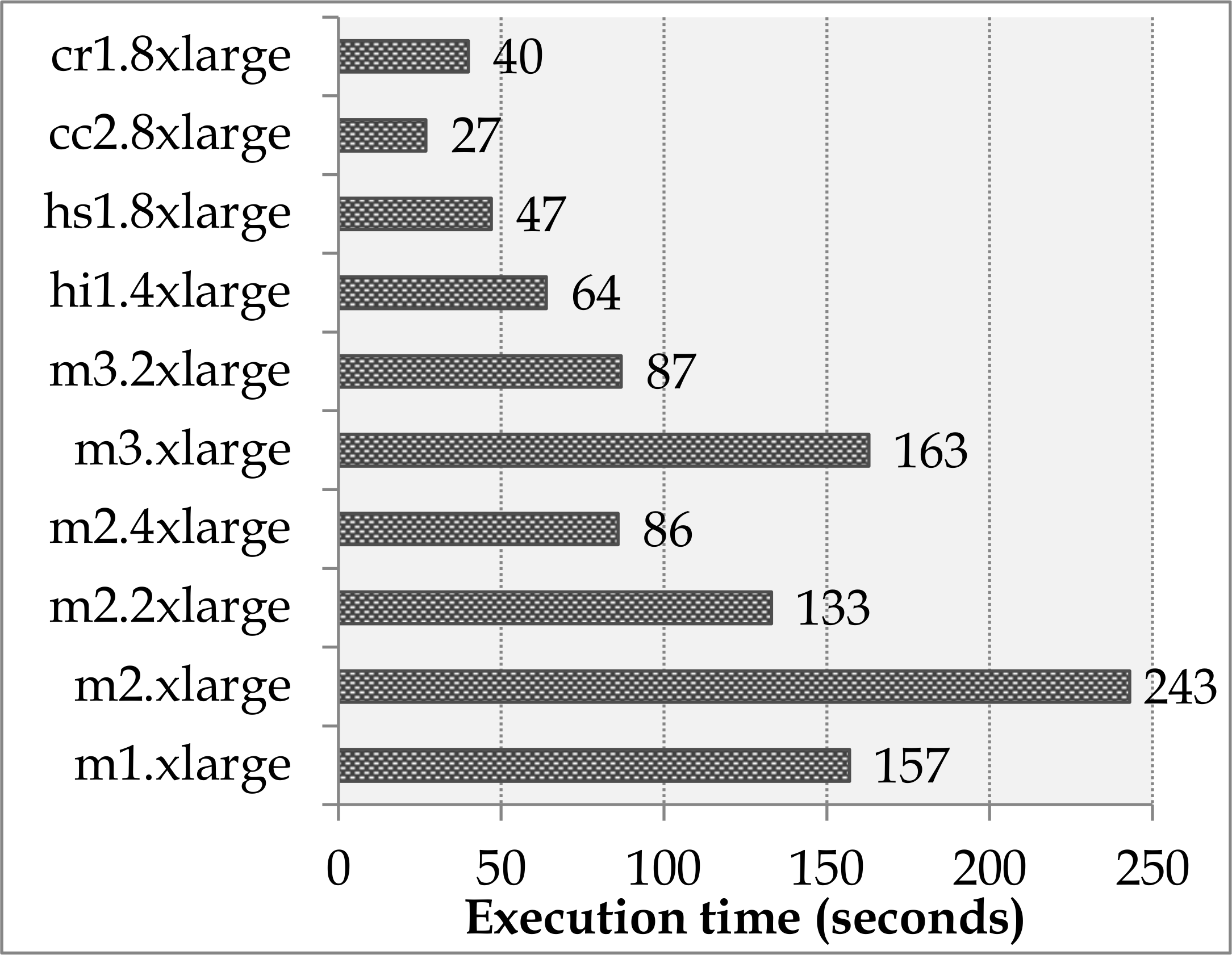}} \hfill
	\subfloat[Case study 3 - parallel]{\label{figure2-3}\includegraphics[width=0.32\textwidth]{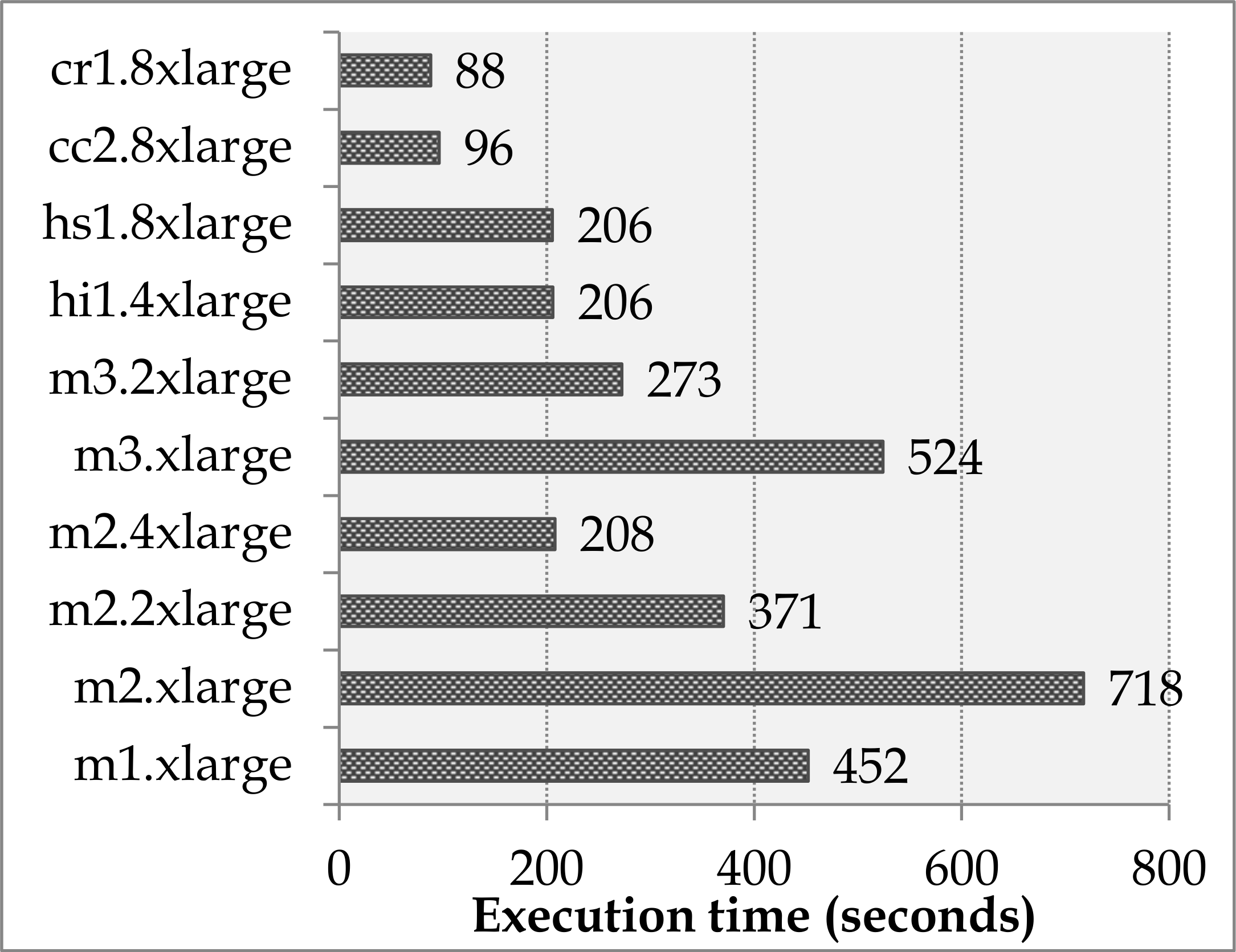}}
\caption{Sequential and parallel execution times for the case study applications}
\label{figure11}
\end{figure*}

\textit{Step 1 - Execute the three case study application on all VMs.}  The time taken to execute the applications sequentially is presented in Figure \ref{figure1-1} to Figure \ref{figure1-3} and to execute the applications in parallel using all available vCPUs is presented in Figure \ref{figure2-1} to Figure \ref{figure2-2}. In all case studies, the \texttt{cr1.8xlarge} and \texttt{cc2.8xlarge} have best performance; these VMs show good performance in memory and process and computation groups. The \texttt{m3} VMs are close competitors for sequential execution and \texttt{hi1.4xlarge} and \texttt{hs1.8xlarge} perform well for parallel execution. The results from parallel execution depend on the number of vCPUs available on the VM.

\textit{Step 2 - Generate empirical ranks for the case study.} The empirical ranks are generated using the standard competition ranking approach. The lowest time translates to the highest rank. If there are two VMs with the same program execution time they get the same rank and the ranking is continued with a gap. For example, in Figure \ref{figure1-1}, \texttt{m3.2xlarge} and \texttt{m3.xlarge} have the same program execution time. Both VMs have third rank and the next best performing VM, \texttt{hs1.8xlarge} obtains the fifth rank. 

\textit{Step 3 - Provide weights of the application to DocLite.} To generate the rankings from DocLite, a user provides the set of weights $W$ that characterise the case study applications which was considered in Section \ref{lightweight}. In consultation with domain scientists and practitioners, the weights for the three case studies are $\{4, 3, 5, 0\}$, $\{5, 3, 5, 0\}$ and $\{2, 0, 5, 0\}$ respectively. The above ranks were provided as input to the two benchmarking methods. 


\textit{Step 4 - Obtain benchmark ranks for the application.} The empirical and benchmarking ranks for the three case studies using the native and hybrid methods are obtained (Tables \ref{table2-1} to \ref{table2-3} show the ranks for the native approach. Tables \ref{table3-1} to \ref{table3-3} show the ranks for the case studies using the hybrid method  in which data from the native method along with data obtained previously from benchmarking the entire VM were considered). 

\begin{table}
\centering
\begin{tabular}{ |p{1.4cm}|p{0.4cm}|p{0.4cm}|p{0.4cm}|p{0.4cm}|p{0.4cm}|p{0.4cm}|p{0.4cm}|p{0.4cm}|  }
\hline
\multirow{2}{*}{Amazon VM} & \multicolumn{4}{|c|}{Sequential Ranking} &
\multicolumn{4}{|c|}{Parallel Ranking}\\ \cline{2-9}
& Emp-irical & 100 MB & 500 MB & 1000 MB & Emp-irical & 100 MB & 500 MB & 1000 MB \\
\hline
\hline
m1.xlarge   & 9    &10     &10     &10     &9      &10     &10     &10 \\
m2.xlarge   & 7     &4      &4      &5      &10     &8      &8      &8 \\
m2.2xlarge  & 6     &7      &6      &7      &7      &9      &9      &9 \\
m2.4xlarge  & 5     &6      &7      &6      &5      &6      &6      &6 \\
m3.xlarge   & 4     &3      &3      &3      &8      &7      &7      &7 \\
m3.2xlarge  & 3     &5      &5      &5      &6      &4      &3      &4 \\
cr1.8xlarge & 1     &1      &1      &1      &1      &1      &1      &1 \\
cc2.8xlarge & 2     &2      &2      &2      &2      &2      &2      &2 \\
hi1.4xlarge & 8     &8      &8      &8      &3      &3      &4      &3 \\
hs1.8xlarge & 10     &9      &9      &9      &4      &5      &5      &5 \\
\hline
\end{tabular}
\caption{Case Study 1: Empirical and benchmark rankings for the native benchmarking method}
\label{table2-1}
\end{table}

\begin{table}
\centering
\begin{tabular}{ |p{1.4cm}|p{0.4cm}|p{0.4cm}|p{0.4cm}|p{0.4cm}|p{0.4cm}|p{0.4cm}|p{0.4cm}|p{0.4cm}|  }
\hline
\multirow{2}{*}{Amazon VM} & \multicolumn{4}{|c|}{Sequential Ranking} &
\multicolumn{4}{|c|}{Parallel Ranking}\\ \cline{2-9}
& Emp-irical & 100 MB & 500 MB & 1000 MB & Emp-irical & 100 MB & 500 MB & 1000 MB \\
\hline
\hline
m1.xlarge 	&	10	&	10	&	10	&	10	&	8	&	10	&	10	&	10\\
m2.xlarge 	&	6	&	5	&	5	&	4	&	10	&	8	&	8	&	8\\
m2.2xlarge 	&	6	&	7	&	6	&	7	&	7	&	9	&	9	&	9\\
m2.4xlarge 	&	6	&	6	&	7	&	6	&	5	&	6	&	6	&	6\\
m3.xlarge 	&	3	&	3	&	3	&	3	&	9	&	7	&	7	&	7\\
m3.2xlarge 	&	3	&	4	&	4	&	5	&	6	&	4	&	4	&	4\\
cr1.8xlarge 	&	1	&	1	&	1	&	1	&	2	&	1	&	1	&	1\\
cc2.8xlarge 	&	2	&	2	&	2	&	2	&	1	&	2	&	2	&	2\\
hi1.4xlarge 	&	9	&	8	&	8	&	8	&	4	&	3	&	3	&	3\\
hs1.8xlarge 	&	5	&	9	&	9	&	9	&	3	&	5	&	5	&	5\\
\hline
\end{tabular}
\caption{Case Study 2: Empirical and benchmark rankings for the native benchmarking method}
\label{table2-2}
\end{table}

\begin{table}
\centering
\begin{tabular}{ |p{1.4cm}|p{0.4cm}|p{0.4cm}|p{0.4cm}|p{0.4cm}|p{0.4cm}|p{0.4cm}|p{0.4cm}|p{0.4cm}|  }
\hline
\multirow{2}{*}{Amazon VM} & \multicolumn{4}{|c|}{Sequential Ranking} &
\multicolumn{4}{|c|}{Parallel Ranking}\\ \cline{2-9}
& Emp-irical & 100 MB & 500 MB & 1000 MB & Emp-irical & 100 MB & 500 MB & 1000 MB \\
\hline
\hline
m1.xlarge 	&	10	&	10	&	10	&	10	&	8	&	10	&	10	&	10\\
m2.xlarge 	&	6	&	5	&	5	&	5	&	10	&	8	&	8	&	8\\
m2.2xlarge 	&	7	&	7	&	7	&	7	&	7	&	9	&	9	&	9\\
m2.4xlarge 	&	5	&	6	&	6	&	6	&	5	&	6	&	6	&	6\\
m3.xlarge 	&	2	&	3	&	3	&	3	&	9	&	7	&	7	&	7\\
m3.2xlarge 	&	3	&	4	&	4	&	4	&	6	&	5	&	5	&	5\\
cr1.8xlarge 	&	1	&	1	&	1	&	1	&	1	&	2	&	2	&	2\\
cc2.8xlarge 	&	4	&	2	&	2	&	2	&	2	&	1	&	1	&	1\\
hi1.4xlarge 	&	8	&	8	&	8	&	8	&	3	&	3	&	3	&	3\\
hs1.8xlarge 	&	9	&	9	&	9	&	9	&	3	&	4	&	4	&	4\\
\hline
\end{tabular}
\caption{Case Study 3: Empirical and benchmark rankings for the native benchmarking method}
\label{table2-3}
\end{table}

\begin{table}
\centering
\begin{tabular}{ |p{1.4cm}|p{0.4cm}|p{0.4cm}|p{0.4cm}|p{0.4cm}|p{0.4cm}|p{0.4cm}|p{0.4cm}|p{0.4cm}|  }
\hline
\multirow{2}{*}{Amazon VM} & \multicolumn{4}{|c|}{Sequential Ranking} &
\multicolumn{4}{|c|}{Parallel Ranking}\\ \cline{2-9}
& Emp-irical & 100 MB & 500 MB & 1000 MB & Emp-irical & 100 MB & 500 MB & 1000 MB \\
\hline
\hline
m1.xlarge 	&	9	&	10	&	10	&	10	&	9	&	10	&	10	&	10\\
m2.xlarge 	&	7	&	5	&	5	&	5	&	10	&	9	&	9	&	9\\
m2.2xlarge 	&	6	&	7	&	7	&	7	&	7	&	8	&	8	&	8\\
m2.4xlarge 	&	5	&	6	&	6	&	6	&	5	&	6	&	6	&	6\\
m3.xlarge 	&	4	&	3	&	3	&	3	&	8	&	7	&	7	&	7\\
m3.2xlarge 	&	3	&	4	&	4	&	4	&	6	&	4	&	4	&	4\\
cr1.8xlarge 	&	1	&	1	&	1	&	1	&	1	&	1	&	1	&	1\\
cc2.8xlarge 	&	2	&	2	&	2	&	2	&	2	&	2	&	2	&	2\\
hi1.4xlarge 	&	8	&	8	&	8	&	8	&	3	&	3	&	3	&	3\\
hs1.8xlarge 	&	10	&	9	&	9	&	9	&	4	&	5	&	5	&	5\\
\hline
\end{tabular}
\caption{Case Study 1: Empirical and benchmark rankings for the hybrid benchmarking method}
\label{table3-1}
\end{table}

\begin{table}
\centering
\begin{tabular}{ |p{1.4cm}|p{0.4cm}|p{0.4cm}|p{0.4cm}|p{0.4cm}|p{0.4cm}|p{0.4cm}|p{0.4cm}|p{0.4cm}|  }
\hline
\multirow{2}{*}{Amazon VM} & \multicolumn{4}{|c|}{Sequential Ranking} &
\multicolumn{4}{|c|}{Parallel Ranking}\\ \cline{2-9}
& Emp-irical & 100 MB & 500 MB & 1000 MB & Emp-irical & 100 MB & 500 MB & 1000 MB \\
\hline
\hline
m1.xlarge 	&	10	&	10	&	10	&	10	&	8	&	10	&	10	&	10\\
m2.xlarge 	&	6	&	5	&	5	&	5	&	10	&	9	&	9	&	9\\
m2.2xlarge 	&	6	&	7	&	7	&	7	&	7	&	8	&	8	&	8\\
m2.4xlarge 	&	6	&	6	&	6	&	6	&	5	&	6	&	6	&	6\\
m3.xlarge 	&	3	&	3	&	3	&	3	&	9	&	7	&	7	&	7\\
m3.2xlarge 	&	3	&	4	&	4	&	4	&	6	&	4	&	4	&	4\\
cr1.8xlarge 	&	1	&	1	&	1	&	1	&	2	&	1	&	1	&	1\\
cc2.8xlarge 	&	2	&	2	&	2	&	2	&	1	&	2	&	2	&	2\\
hi1.4xlarge 	&	9	&	8	&	8	&	8	&	4	&	3	&	3	&	3\\
hs1.8xlarge 	&	5	&	9	&	9	&	9	&	3	&	5	&	5	&	5\\
\hline
\end{tabular}
\caption{Case Study 2: Empirical and benchmark rankings for the hybrid benchmarking method}
\label{table3-2}
\end{table}

\begin{table}
\begin{tabular}{ |p{1.4cm}|p{0.4cm}|p{0.4cm}|p{0.4cm}|p{0.4cm}|p{0.4cm}|p{0.4cm}|p{0.4cm}|p{0.4cm}|  }
\hline
\multirow{2}{*}{Amazon VM} & \multicolumn{4}{|c|}{Sequential Ranking} &
\multicolumn{4}{|c|}{Parallel Ranking}\\ \cline{2-9}
& Emp-irical & 100 MB & 500 MB & 1000 MB & Emp-irical & 100 MB & 500 MB & 1000 MB \\
\hline
\hline
m1.xlarge 	&	10	&	10	&	10	&	10	&	8	&	10	&	10	&	10\\
m2.xlarge 	&	6	&	5	&	5	&	5	&	10	&	9	&	9	&	9\\
m2.2xlarge 	&	7	&	7	&	7	&	7	&	7	&	8	&	8	&	8\\
m2.4xlarge 	&	5	&	6	&	6	&	6	&	5	&	6	&	6	&	6\\
m3.xlarge 	&	2	&	3	&	3	&	3	&	9	&	7	&	7	&	7\\
m3.2xlarge 	&	3	&	4	&	4	&	4	&	6	&	4	&	4	&	4\\
cr1.8xlarge 	&	1	&	1	&	1	&	1	&	1	&	1	&	1	&	1\\
cc2.8xlarge 	&	4	&	2	&	2	&	2	&	2	&	2	&	2	&	2\\
hi1.4xlarge 	&	8	&	8	&	8	&	8	&	3	&	3	&	3	&	3\\
hs1.8xlarge 	&	9	&	9	&	9	&	9	&	3	&	5	&	5	&	5\\
\hline
\end{tabular}
\caption{Case Study 3: Empirical and benchmark rankings for the hybrid benchmarking method}
\label{table3-3}
\end{table}

Sequential and parallel ranks are generated for each case study using the weights. The empirical ranks are obtained from the timing results. The ranks obtained when using different sizes of the container are also reported in the tables. 

As mentioned in Section \ref{benchmarking}, if there are $i=1, 2, \cdots, m$ different VMs, and each VM is represented as $vm_{i}$, then the performance rank obtained from each VM is $Rp_{i}$. If the empirical rank of each VM by executing the case study application is $Re_{i}$, then the absolute distance between the ranks is $d = |Rp_{i} - Re_{i}|$. The sum of distances can be represented as $d_{s} = \sum\limits_{i=1}^m |Rp_{i} - Re_{i}|$. 

\begin{figure*}[ht]
\centering
	\subfloat[Sequential execution]{\label{figure55-1}\includegraphics[width=0.48\textwidth]{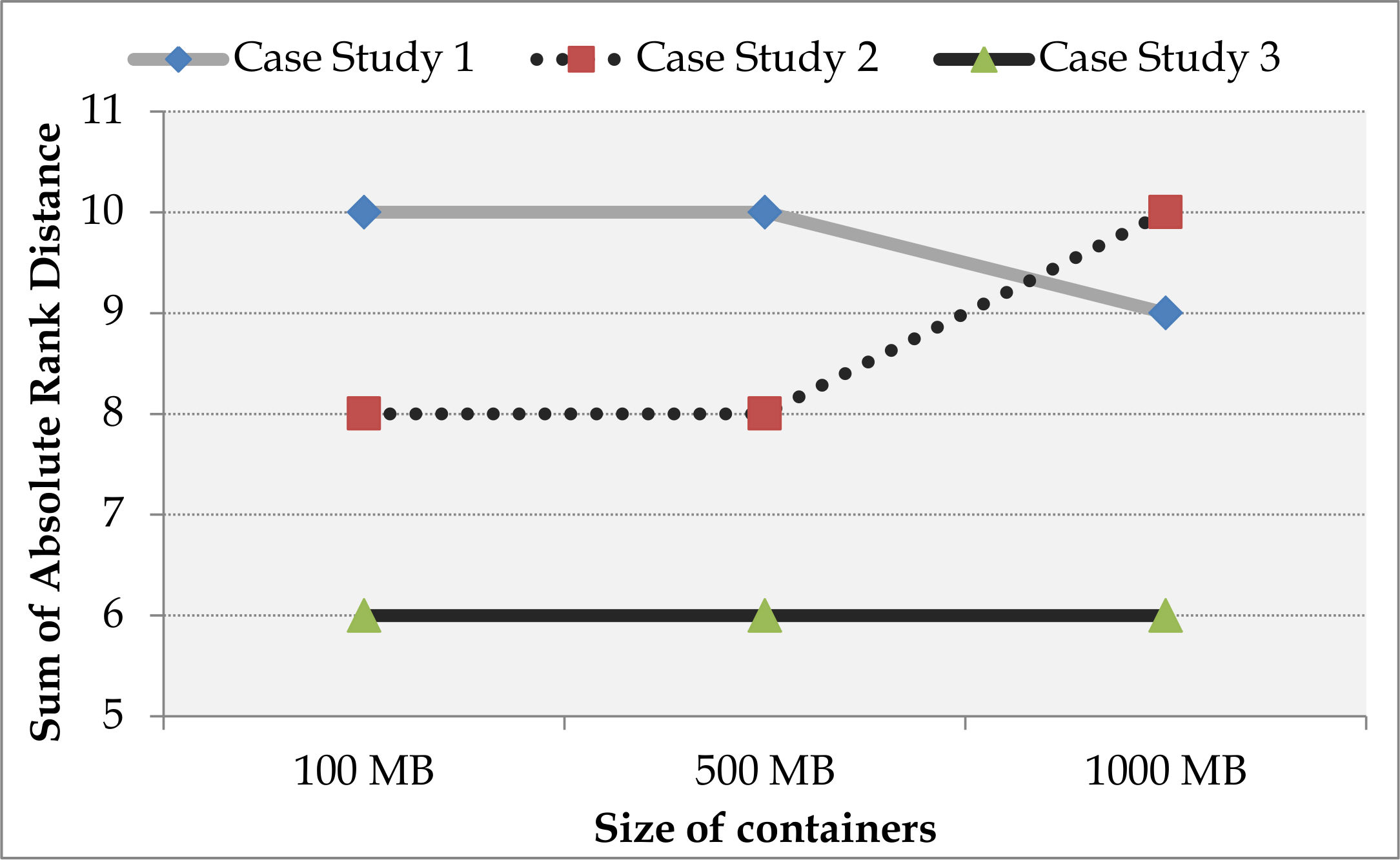}} \hfill
	\subfloat[Parallel execution]{\label{figure55-2}\includegraphics[width=0.48\textwidth]{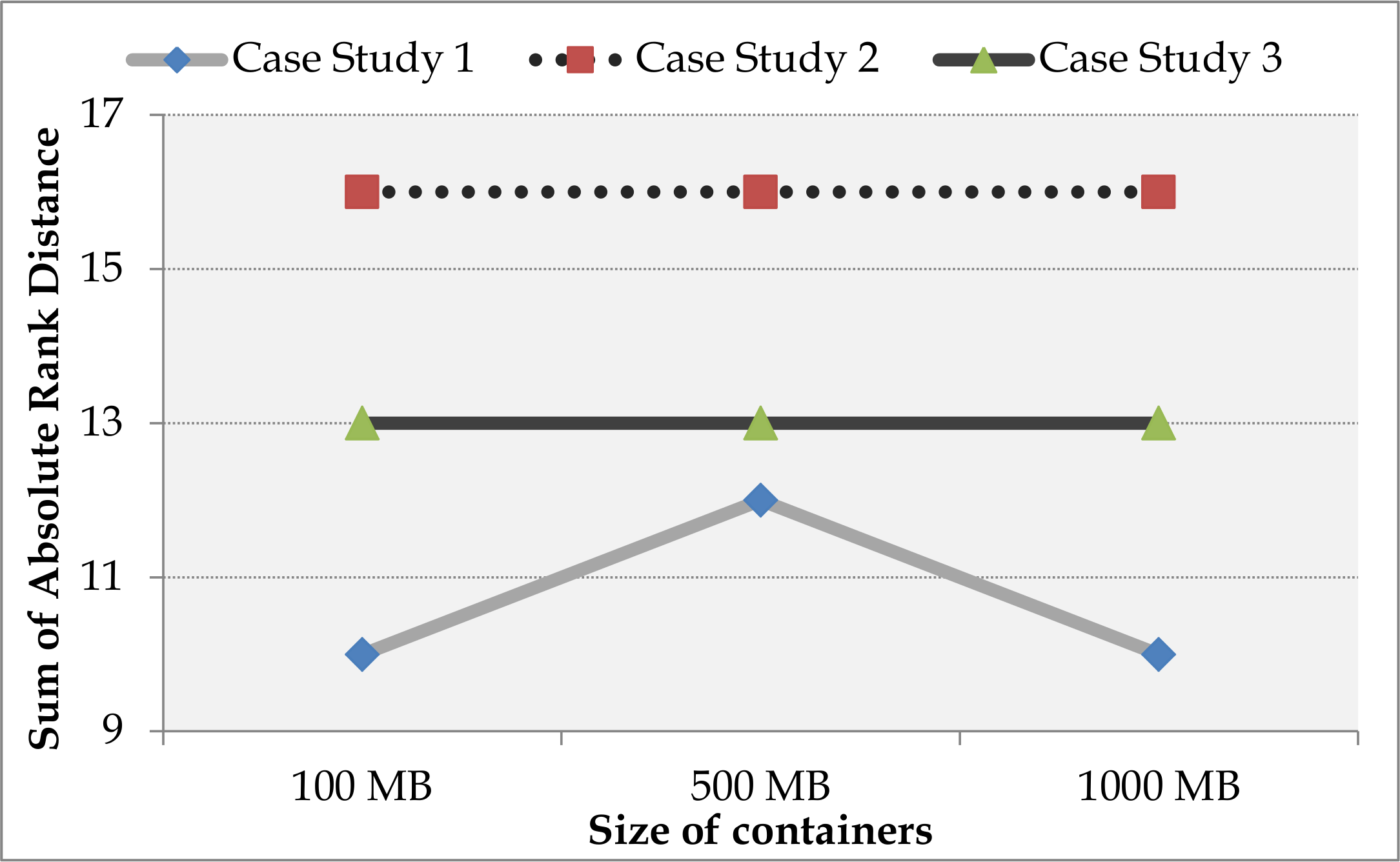}} 
\caption{Sum of absolute distances for the native benchmarking method}
\label{figure55}
\end{figure*}

\begin{figure*}[ht]
\centering
	\subfloat[Sequential execution]{\label{figure66-1}\includegraphics[width=0.48\textwidth]{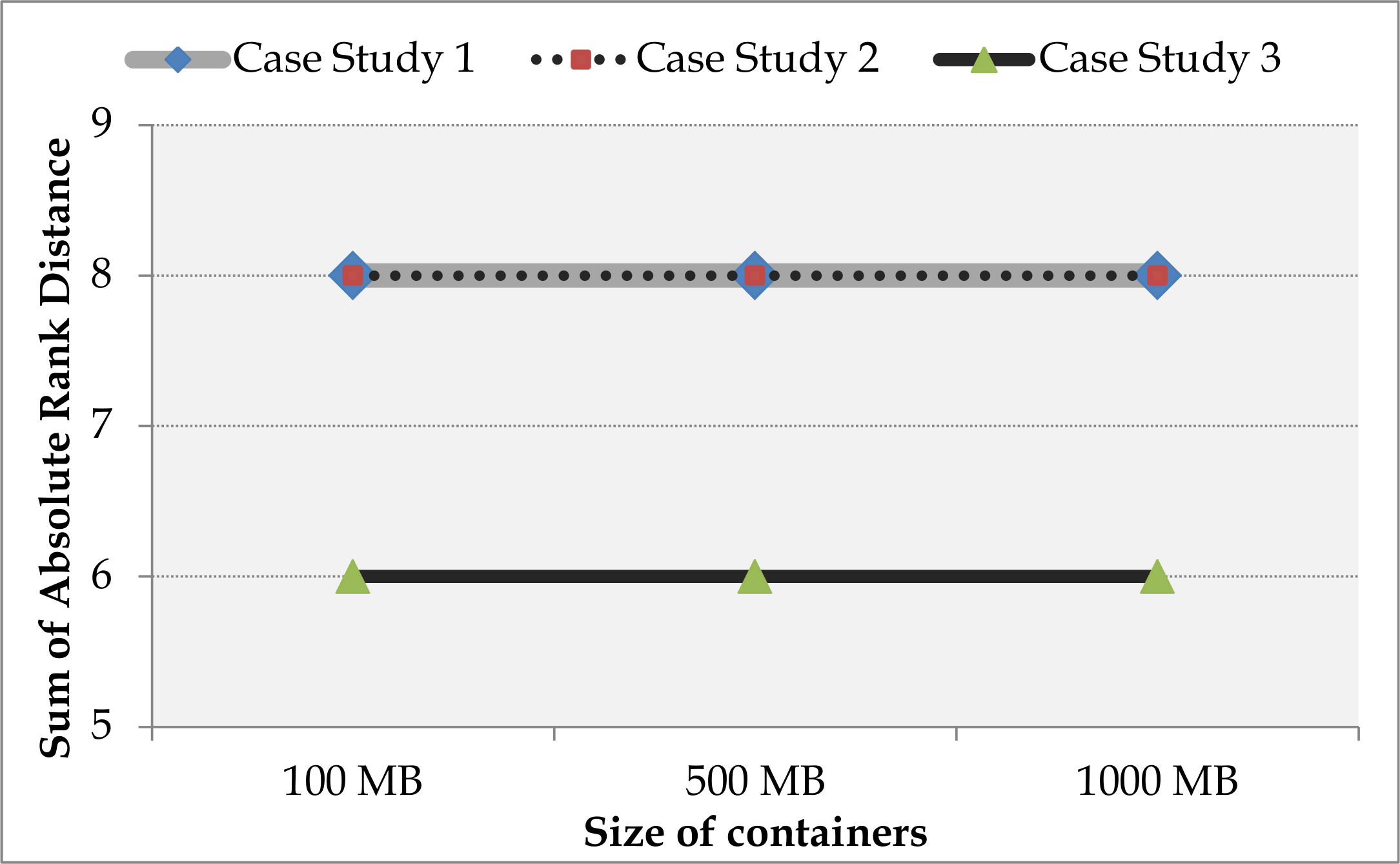}} \hfill
	\subfloat[Parallel execution]{\label{figure66-2}\includegraphics[width=0.48\textwidth]{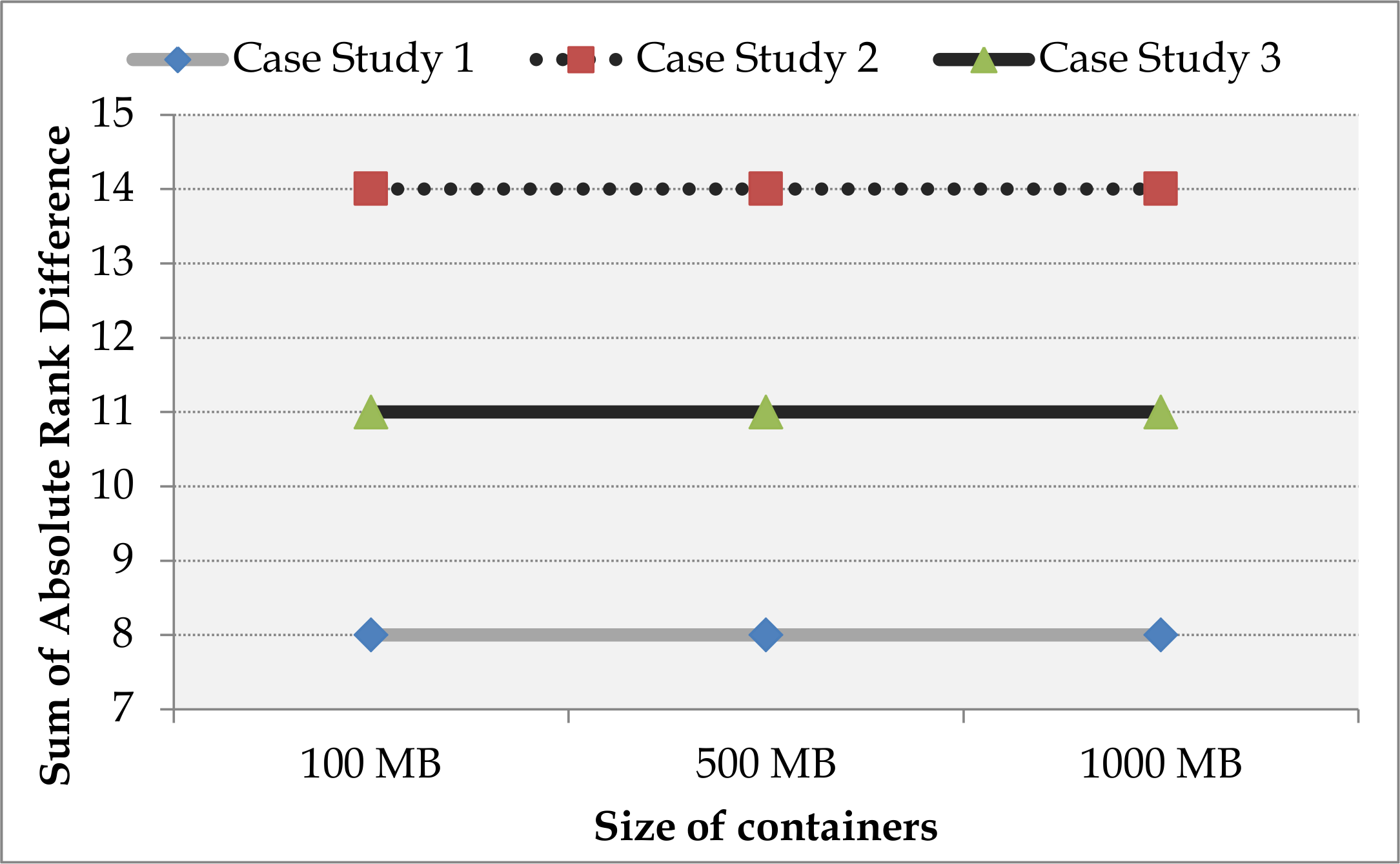}} 
\caption{Sum of absolute distances for the hybrid benchmarking method}
\label{figure66}
\end{figure*}

Figure \ref{figure55} and Figure \ref{figure66} shows the sum of rank distances of sequential and parallel executions of each case study using 100 MB, 500 MB and 1000 MB containers using the native and hybrid methods respectively. The sum of rank distances provide one view for comparing the ranks obtained from using different container sizes; lower values of the sum of rank distances translate to a higher correlation between empirical and benchmark ranks.  
Consider the plot of sequential execution of Case Study 3 for example in Figure \ref{figure55-1}. It is inferred that using a large container does not reduce the sum of rank distances. Different size of containers produce the same sum of rank distances. With the exception of sequential execution for Case Study 1 in Figure \ref{figure55-1} and Figure \ref{figure55-2}, a larger size container does not produce better results than a 100 MB container. This inference is again confirmed by the plots in Figure \ref{figure66}. 

\textit{Step 5 - Find correlation of benchmark and empirical ranks.} Given the rank tables for each case study it is important to determine the accuracy (or quality) of the ranks. In this paper, the accuracy of results is the correlation between the empirical ranks and the benchmark ranks. This quality measure validates the feasibility of using lightweight benchmarks and guarantees results obtained from benchmarking correspond to reality. 

\begin{table}[h]
\centering
\begin{tabular}{ |c|p{0.6cm}|p{0.6cm}|p{0.6cm}|p{0.6cm}|p{0.6cm}|p{0.6cm}|p{0.6cm}|p{0.6cm}|  }
\hline
\multirow{2}{*}{Case study} & \multicolumn{3}{|c|}{Sequential Ranking} &
\multicolumn{3}{|c|}{Parallel Ranking}\\ \cline{2-7}
& 100 MB & 500 MB & 1000 MB & 100 MB & 500 MB & 1000 MB \\
\hline
\hline
\multicolumn{7}{|l|}{Native method}\\
\hline
1 & 89.1 & 87.9 & 92.1 & 90.3 & 86.7 & 90.3\\
2 & 88.5 & 88.5 & 84.7 & 83.0 & 83.0 & 83.0\\
3 & 95.2 & 95.2 & 95.2 & 87.6 & 87.6 & 87.6\\
\hline
\multicolumn{7}{|l|}{Hybrid method}\\
\hline
1 & 93.9 & 93.9 & 93.9 & 93.9 & 93.9 & 93.9\\
2 & 88.5 & 88.5 & 88.5 & 86.7 & 86.7 & 86.7\\
3 & 95.2 & 95.2 & 95.2 & 88.8 & 88.8 & 88.8\\
\hline
\end{tabular}
\caption{Correlation (in \%) between empirical and benchmarking ranks for the native and hybrid benchmarking methods}
\label{table4-1}
\end{table}


In \textit{Step 5}, the correlation of the benchmark ranks using different containers and the empirical ranks for benchmarking is determined and shown in Table \ref{table4-1}; the percentage value shows the degree of correlation. Higher the correlation value the more robust is the benchmarking method since it corresponds more closely to the empirical ranks.

There is on an average there is over 90\% and 86\% correlation between the empirical and benchmarked sequential and parallel ranks respectively for the native method. It is observed that increasing the size of the container does not generally increase the correlation between the ranks. The smallest container of 100 MB performs as well as the other containers. There is an average improvement of 1\%-2\% in the correlation between the ranks for the hybrid method. While the hybrid method can improve the ranks, it is observed that the position of the top three ranks are not affected. Again, using smaller containers do not change the quality of results.   

\subsection{Summary} 
Three key observations are summarised from the experimental studies. Firstly, small containers using lightweight benchmarks perform similar to large containers. There is no significant improvement in the quality of results with larger containers. On average, there is over 90\% and 86\% correlation when comparing ranks obtained from the empirical analysis and the 100 MB container.

Secondly, the hybrid method slightly improves the quality of the benchmark rankings, although the position of the top three ranks do not change. The lightweight method is sufficient to maximise the performance of an application on the cloud. Implementing hybrid methods will require the storage of historic benchmark data and its maintenance over time. Efficient methods for assigning weights to data based on how recent it is will need to be developed. 

Thirdly, container-based benchmarks have significantly lower execution time when compared to executing benchmarks directly on the VM. This highlights the use of lightweight methods for real-time deployment of applications.

\section{Related Work}
\label{relatedwork}
Benchmarking can capture the performance of a computing system \cite{rel-1}. Standard benchmarks such as Linpack are used for ranking the top supercomputers \cite{rel-2}. However, there are fewer standards for benchmarking methods that are accepted by the cloud community. Currently, there are a number of ongoing efforts to develop standard methods applicable on the cloud \cite{cloudbenchmark-4, rel-5}. Benchmarking is usually performed directly on a VM by using all resources available to the entire VM. This is done in order to generate accurate benchmarks and takes a few hours to complete on large VMs. Such time consuming benchmarking methods cannot be used in real-time although they can generate accurate and detailed benchmark data.

Benchmarking methods need to be (i) employed in near real-time and (ii) produce accurate benchmarks to facilitate cloud performance benchmarking in a meaningful way. This is important because VMs have different performance characteristics over time and sometimes even during small time periods. For example, a real-time benchmarking method is used for selecting VMs that match application requirements by Netflix \cite{foot-1}. Alternate virtualisation technology, such as containers with low boot up times and a high degree of resource isolation are likely to be the way forward to achieve lightweight methods that are not only fast but also produce reliable benchmark data \cite{cont-2, cont-3}.

There is preliminary research that highlights the lower overheads of containers when compared to existing virtualisation technologies both for both high-performance computing systems as well as for the cloud \cite{rel-6, rel-7}. Containers on the cloud as a research topic is gaining momentum and there is recent research reporting the benefit of containers for distributed storage \cite{rel-8}, reproducibility of research \cite{rel-9}, and in the context of security \cite{rel-10}. However, container technology has not yet been adopted for benchmarking. In this paper, we developed lightweight benchmarking methods for VMs that can be used in near real-time to produce reliable benchmarks. 

\section{Conclusions}
\label{conclusions}
Benchmarking is important for selecting VMs that can maximise the performance of an application on the cloud. However, current benchmarking methods are time consuming since they benchmark an entire VM for obtaining accurate benchmarks, thereby limiting their real-time use. In this paper, we explored an alternative to existing benchmarking methods to generate accurate benchmarks in near real-time by using containers as a means to achieve lightweight benchmarking. 

In the research presented in this paper, Docker Container-based Lightweight Benchmarking tool, referred to as `DocLite' was developed to facilitate lightweight benchmarking. DocLite organises the benchmark data into four groups, namely memory and process, local communication, computation and storage. A user of DocLite provides as input a set of four weights (ranging from 0 to 5), which indicate how important each of the groups are to the application that needs to be deployed on the cloud. The weights are mapped onto the four benchmark groups and are used to generate a score for ranking the VMs according to performance. DocLite incorporates two benchmarking methods. In the first mode, referred to as the native method, containers are used to benchmark a portion of the VM to generate ranks of cloud VMs, and the second in which data obtained from the first method is used in conjunction with historic data as a hybrid. DocLite is available to download from https://github.com/lawansubba/DoCLite. Benchmarking using DocLite is between 19-91 times faster than benchmarking the entire VM making lightweight methods suitable for use in real-time. The experimental results highlight that the benchmarks obtained from container-based methods are on an average over 90\% accurate making them as reliable as benchmarking an entire VM. 
Container-based technology is useful for benchmarking on the cloud and can be used for developing fast and reliable benchmarking methods. 

The native and hybrid approaches only considers local communication and do not take network communication into account. In the future, we aim to extend the tool for including network communications to benchmark applications that execute on multiple VMs and data centres.

\section*{Acknowledgment}
This research was pursued under the EPSRC grant, EP/K015745/1, a Royal Society Industry Fellowship, an Erasmus Mundus Master's scholarship and an AWS Education Research grant.

\balance


\end{document}